\documentclass[12pt]{article}

\usepackage{jheppub} 
\usepackage{float}
\usepackage{amsmath,bbm,blkarray,bm}
\usepackage[vcentermath]{youngtab}
\usepackage{slashed}
\usepackage{bbold}
\usepackage{mathtools}
\usepackage{booktabs}
\usepackage{multirow}
\usepackage{tikz}
\usepackage{changepage}
\usepackage{makecell}
\usepackage{tikz-feynman}
\usepackage{subcaption}
\usepackage{color}
\usepackage[numbers,sort&compress]{natbib}
\usepackage[T1]{fontenc} 
\usepackage{amsmath}
\usepackage{bbm}
\usepackage{subcaption}

\definecolor{maroon}{cmyk}{0, 0.87, 0.68, 0.32}
\definecolor{halfgray}{gray}{0.55}
\definecolor{ipython_frame}{RGB}{207, 207, 207}
\definecolor{ipython_bg}{RGB}{247, 247, 247}
\definecolor{ipython_red}{RGB}{186, 33, 33}
\definecolor{ipython_green}{RGB}{0, 128, 0}
\definecolor{ipython_cyan}{RGB}{64, 128, 128}
\definecolor{ipython_purple}{RGB}{170, 34, 255}

\usepackage{listings}
\lstdefinelanguage{iPython}{
    morekeywords={access,and,break,class,continue,def,del,elif,else,except,exec,finally,for,from,global,if,import,in,is,lambda,not,or,pass,print,raise,return,try,while},%
    morekeywords=[2]{abs,all,any,basestring,bin,bool,bytearray,callable,chr,classmethod,cmp,compile,complex,delattr,dict,dir,divmod,enumerate,eval,execfile,file,filter,float,format,frozenset,getattr,globals,hasattr,hash,help,hex,id,input,int,isinstance,issubclass,iter,len,list,locals,long,map,max,memoryview,min,next,object,oct,open,ord,pow,property,range,raw_input,reduce,reload,repr,reversed,round,set,setattr,slice,sorted,staticmethod,str,sum,super,tuple,type,unichr,unicode,vars,xrange,zip,apply,buffer,coerce,intern},%
    sensitive=true,%
    morecomment=[l]\#,%
    morestring=[b]',%
    morestring=[b]",%
    morestring=[s]{'''}{'''},
    morestring=[s]{"""}{"""},
    morestring=[s]{r'}{'},
    morestring=[s]{r"}{"},%
    morestring=[s]{r'''}{'''},%
    morestring=[s]{r"""}{"""},%
    morestring=[s]{u'}{'},
    morestring=[s]{u"}{"},%
    morestring=[s]{u'''}{'''},%
    morestring=[s]{u"""}{"""},%
    literate=
    {á}{{\'a}}1 {é}{{\'e}}1 {í}{{\'i}}1 {ó}{{\'o}}1 {ú}{{\'u}}1
    {Á}{{\'A}}1 {É}{{\'E}}1 {Í}{{\'I}}1 {Ó}{{\'O}}1 {Ú}{{\'U}}1
    {à}{{\`a}}1 {è}{{\`e}}1 {ì}{{\`i}}1 {ò}{{\`o}}1 {ù}{{\`u}}1
    {À}{{\`A}}1 {È}{{\'E}}1 {Ì}{{\`I}}1 {Ò}{{\`O}}1 {Ù}{{\`U}}1
    {ä}{{\"a}}1 {ë}{{\"e}}1 {ï}{{\"i}}1 {ö}{{\"o}}1 {ü}{{\"u}}1
    {Ä}{{\"A}}1 {Ë}{{\"E}}1 {Ï}{{\"I}}1 {Ö}{{\"O}}1 {Ü}{{\"U}}1
    {â}{{\^a}}1 {ê}{{\^e}}1 {î}{{\^i}}1 {ô}{{\^o}}1 {û}{{\^u}}1
    {Â}{{\^A}}1 {Ê}{{\^E}}1 {Î}{{\^I}}1 {Ô}{{\^O}}1 {Û}{{\^U}}1
    {œ}{{\oe}}1 {Œ}{{\OE}}1 {æ}{{\ae}}1 {Æ}{{\AE}}1 {ß}{{\ss}}1
    {ç}{{\c c}}1 {Ç}{{\c C}}1 {ø}{{\o}}1 {å}{{\r a}}1 {Å}{{\r A}}1
    {€}{{\EUR}}1 {£}{{\pounds}}1
    {^}{{{\color{ipython_purple}\^{}}}}1
    {=}{{{\color{ipython_purple}=}}}1
    {+}{{{\color{ipython_purple}+}}}1
    {*}{{{\color{ipython_purple}$^\ast$}}}1
    {/}{{{\color{ipython_purple}/}}}1
    {+=}{{{+=}}}1
    {-=}{{{-=}}}1
    {*=}{{{$^\ast$=}}}1
    {/=}{{{/=}}}1,
    literate=
    *{-}{{{\color{ipython_purple}-}}}1
     {?}{{{\color{ipython_purple}?}}}1,
    identifierstyle=\color{black}\ttfamily,
    commentstyle=\color{ipython_cyan}\ttfamily,
    stringstyle=\color{ipython_red}\ttfamily,
    keepspaces=true,
    showspaces=false,
    showstringspaces=false,
    rulecolor=\color{ipython_frame},
    frame=single,
    frameround={t}{t}{t}{t},
    framexleftmargin=6mm,
    numbers=left,
    numberstyle=\tiny\color{halfgray},
    backgroundcolor=\color{ipython_bg},
    basicstyle=\footnotesize\ttfamily,
    keywordstyle=\color{ipython_green}\ttfamily,
    aboveskip=1.2em,
    belowskip=1.2em,
}

\newcommand{\wilson}{\texttt{wilson}}
\newcommand{\WCxf}{\texttt{WCxf}}

\usepackage{adjustbox}
\usepackage{cancel}

\newcommand{\Q}{\mathcal {O}}
\newcommand{\C}{\mathcal {C}}

\newcommand{\beq}{\begin{equation}}
\newcommand{\eeq}{\end{equation}}
\newcommand{\be}{\begin{equation}}
\newcommand{\ee}{\end{equation}}
\newcommand{\bi}{\begin{itemize}}
\newcommand{\ei}{\end{itemize}}
\newcommand{\ba}{\begin{array}}
\newcommand{\ea}{\end{array}}
\newcommand{\beqa}{\begin{eqnarray}}
\newcommand{\eeqa}{\end{eqnarray}}
\newcommand{\bea}{\begin{eqnarray}}
\newcommand{\eea}{\end{eqnarray}}
\newcommand{\beqn}{\begin{eqnarray}}
\newcommand{\eeqn}{\end{eqnarray}}

\definecolor{red}{cmyk}{0,1,1,0.4}

\usepackage{slashed}

\newcommand{\wc}[3][{}]{{C_{\! \underset{ #3}{ #2}}^{{\tiny #1}}} }

\allowdisplaybreaks

\title{\includegraphics[width=4cm]{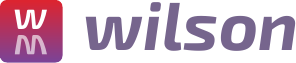} \\
A package for renormalization group running in the SMEFT with Sterile Neutrinos}

\author[a]{Jason Aebischer,}
\author[b]{Tejhas Kapoor,}
\author[c]{Jacky Kumar}

\affiliation[a]{Theoretical Physics Department, CERN, 1211 Geneva 23, Switzerland}
\affiliation[b]{Universit\'e Paris-Saclay, CNRS/IN2P3, IJCLab, 91405 Orsay, France}
\affiliation[c]{Theoretical Division, Los Alamos National Laboratory, Los Alamos, NM 87545, USA}

\emailAdd{jason.aebischer@cern.ch}
\emailAdd{jacky.kumar@lanl.gov}
\emailAdd{tejhas.kapoor@etu-upsaclay.fr}

\begin{document}
\begin{flushright}
{
CERN-TH-2024-186\\
LA-UR-24-31845
}
\end{flushright}

\abstract{Sterile neutrinos are well-motivated beyond the Standard Model (BSM) particles. The Standard Model Effective Field Theory (SMEFT) augmented with these new fields is known as the $\nu$SMEFT. We present the first code for solving the renormalization group equations (RGEs) of the $\nu$SMEFT in an automated way. For this purpose, we have implemented the $\nu$SMEFT as a new effective field theory (EFT) in the Wilson coefficient exchange format \WCxf. Furthermore, we included anomalous dimensions depending on the gauge couplings and Yukawas in the python package \wilson{}\footnote{\url{https://wilson-eft.github.io/}}. This novel version of \wilson{} allows a consistent inclusion of $\nu$SMEFT renormalization group (RG) running effects above the electroweak (EW) scale in phenomenological studies involving sterile neutrinos. Moreover, this new release allows us to study EW, strong, and Yukawa running effects separately within the SMEFT. }

\maketitle

\thispagestyle{empty}
\newpage

\section{Introduction}
Right-handed neutrinos are a natural extension of the Standard Model (SM) of particle physics. Since the observation of neutrino oscillations, neutrinos are known to have non-zero masses. The mass hierarchy remains however yet to be determined \cite{Esteban:2024eli}, exhibiting either normal ordering, i.e. $m_1 < m_2 < m_3$ or inverse ordering ($m_3 < m_1 < m_2$). In order for the neutrinos to become mass states, a simple solution is to add right-handed neutrinos to the SM. {These can acquire a Majorana mass or, in combination with the coupling to the Higgs a Dirac mass.} Allowing for even higher-dimensional operators involving right-handed neutrinos and the SM field content one arrives at the so-called $\nu$SMEFT, which is the Standard Model Effective Field Theory \cite{Buchmuller:1985jz} (for reviews, see  \cite{Brivio:2017vri,Isidori:2023pyp}), augmented by right-handed neutrinos. 

\par
The $\nu$SMEFT has been studied extensively in the past few years \cite{delAguila:2008ir,Aparici:2009fh,Bhattacharya:2015vja,Liao:2016qyd,Bischer:2019ttk,Alcaide:2019pnf, Han:2022uho,Datta:2022czw}.
However, concerning  loop corrections, the $\nu$SMEFT is not as mature as SMEFT. Recently 
the renormalization of the $\nu$SMEFT has been completed at the 1-loop level \cite{Chala:2020pbn, Datta:2020ocb, Datta:2021akg, Fuyuto:2024oii,Ardu:2024tzb}. 
The one-loop anomalous dimensions of the $\nu$SMEFT depending upon both the gauge and Yukawa couplings have been computed in these references.  Such effects can be important for the running between the new physics (NP) scale $\Lambda$ and the electroweak (EW) scale. Indeed these turn out to be crucial for phenomenological studies in the $\nu$SMEFT \cite{Chala:2020pbn, Cirigliano:2021peb, Fuyuto:2024oii}.  However, as of now, there is no public code that provides numerical solutions to the RG running within the $\nu$SMEFT for the complete set of operators. 

\par 
In this work, we present a major upgrade of \wilson{} \cite{Aebischer:2018bkb}, a Python package for the running and matching of Wilson coefficients (WCs) above and below the EW scale. Provided with the numerical values of the WCs at a high NP scale, the original \wilson{} package is capable of performing the RG evolution within the SMEFT \cite{Jenkins:2013zja,Jenkins:2013wua,Alonso:2013hga}, matching onto the weak effective theory (WET) at the EW scale \cite{Aebischer:2015fzz,Jenkins:2017jig,Dekens:2019ept}, as well as to perform the full QCD/QED RG evolution below the EW scale down to hadronic scales relevant for low-energy precision tests \cite{Aebischer:2017gaw,Jenkins:2017dyc}. In this upgraded version, we have included the functionality of solving the renormalization group equations (RGEs) of the $\nu$SMEFT in an automated way. Moreover, the subtle non-standard RG running effects due to back-rotation \cite{Aebischer:2020lsx} can now be included in $\nu$SMEFT.

\par 
The article is organized as follows. In Section~\ref{sec:nusmeft}, we introduce the $\nu$SMEFT, including its Lagrangian together with the $ \rm SU(3)^6$ flavour rotations of the $\nu$SMEFT operators. In Section~\ref{sec:evolution}, we discuss the RG running in the $\nu$SMEFT. In Section~\ref{sec:nusmeft_wilson}, the implementation of the $\nu$SMEFT in \wilson{} is discussed, along with additional upgrades. In Section~\ref{sec:comparison}, we provide details of the checks performed on the output of our code, and in Section~\ref{sec:summary} we give a summary and future prospects.

{\boldmath
\section{$\nu$SMEFT}\label{sec:nusmeft}
}
\noindent
In this section, we discuss our conventions regarding the $\nu$SMEFT and choice of flavour basis for $\nu$SMEFT operators involving fermions.

\subsection{Lagrangian and Operator Basis}

In addition to the SM fields, the $\nu$SMEFT contains 
sterile neutrinos. We denote the corresponding fields by $n_p$, where $p$ is the generation index and we assume $p \in \{1,2,3 \}$. 

The $\nu$SMEFT Lagrangian is given by
\begin{equation} \label{eq:nuSMEFTLag}
{\cal L}_{\nu {\rm SMEFT}}  \supset  i\bar{n} \slashed{\partial}n+ {(-\frac{1}{2}m_\nu( n^TCn)+h.c.)}+ {\cal L}_{\rm{Yukawa}}+ (\sum_{i} \mathcal{C}_i \Q_i + h.c.)\,,
\end{equation}
where the first term is the kinetic term for the sterile neutrinos. If neutrinos are assumed to be of Majorana nature, one can also add a Majorana mass term $m_\nu$. The presence of such a term does not explicitly enter most of the anomalous dimension matrices (ADMs). However, one exception is the mixing of dipole operators into the Weinberg operator \cite{Fuyuto:2024oii}.  In the current implementation, this piece of the ADM is not included. In the Majorana mass term $C$ stands for the charge-conjugate operator. Furthermore, $\mathcal{C}_i$ are the WCs of the higher dimension ($\ge 5$) operators. We have omitted the ${\cal L}_{\rm SM}$ part containing the usual dimension-four SM terms. Note that our actual implementation of the $\nu$SMEFT in \wilson{} follows \WCxf{} conventions \cite{Aebischer:2017ugx}, in which the complex conjugated part for the higher dimensional operators is added only for the non-hermitian operators. In this way, only one out of two operators related through hermitian conjugation is considered in the basis.

A subset of ${\cal L}_{\rm SM}$, the Yukawa terms plus a new Dirac mass term for the neutrinos are given by
\beq \label{eq:Yuklag}
{\cal L}_{\rm Yukawa} = -[\phi^{\dagger j}\bar{d} Y_d q_j + \tilde{\phi}^{\dagger j}\bar{u} Y_u q_j + \phi^{\dagger j}\bar{e} Y_e \ell_j + \tilde{\phi}^{\dagger j}\bar{n} Y_n \ell_j + \text{h.c.}]~,
\eeq
where $\phi$ is the Higgs doublet and its conjugate field is $\tilde{\phi}^j = \epsilon^{jk} \phi_k^*$. Note that the neutrino Yukawa matrix $Y_n$ is a NP parameter, unlike the other dimension-four terms in ${\cal L}_{\rm SM}$.  

Ignoring the different flavour permutations, there are in total 16 ($\Delta B = 0,  \Delta L =0$) new dimension-six operators as compared to the SMEFT. We assume that the right-handed neutrinos have a lepton number $L=1$. In Table~\ref{tab:SMEFTops1}, we show these new operators in the Warsaw basis convention \cite{Grzadkowski:2010es}, where $prst$ are the flavour indices.  In addition, we also show two $B-$ and $L-$ violating operators in Table~\ref{tab:SMEFTops4} that were first discussed in \cite{Alonso:2014zka} and agree with the findings in \cite{Aebischer:2022wnl}. Further, two new dimension-five operators are given in Table~\ref{tab:SMEFTops3}.

The remaining  Warsaw basis operators which are also part of the $\nu$SMEFT can be found in \cite{Grzadkowski:2010es}.   
The complete set of $\nu$SMEFT operators in \WCxf{} conventions with explicit flavour indices will be published on the \href{https://WCxf.github.io/}{\WCxf{} webpage.}

\begin{table}
\centering
\renewcommand{\arraystretch}{1.5}

\begin{subtable}{\textwidth}
\centering
\begin{tabular}{c|c|c|c|c|c}
\hline
\multicolumn{2}{c}{$(\bar RR)(\bar RR)$} &
\multicolumn{2}{c}{$(\bar LL)(\bar RR)$} &
\multicolumn{2}{c}{$(\bar LR)(\bar RL)$ and $(\bar LR)(\bar LR)$ }\\
\hline
$\Q_{nd}$  & $(\bar n_p \gamma_\mu n_r)(\bar d_s \gamma^\mu d_t)$ &
$\Q_{qn}$  & $(\bar q_p \gamma_\mu q_r)(\bar n_s \gamma^\mu n_t)$ &
$\Q_{\ell n \ell e}$  & $(\bar \ell^j_p  n_r)\epsilon_{jk}(\bar \ell^k_s e_t)$   \\
$\Q_{nu}$  & $(\bar n_p \gamma_\mu  n_r)(\bar u_s \gamma^\mu  u_t)$ &
$\Q_{\ell n}$ & $(\bar \ell_p \gamma_\mu \ell_r)(\bar n_s\gamma^\mu n_t)$ &
$\Q_{\ell n q d}^{(1)}$ &  $(\bar \ell^j_p  n_r)\epsilon_{jk}(\bar q^k_s  d_t)$  \\
$\Q_{ne}$  & $(\bar n_p \gamma_\mu n_r)(\bar e_s \gamma^\mu e_t)$ &
& &
$\Q_{\ell n q d}^{(3)}$  & $(\bar \ell^j_p  \sigma_{\mu\nu} n_r)\epsilon_{jk}(\bar q^k_s \sigma^{\mu\nu} d_t)$     \\
$\Q_{nn}$  & $(\bar n_p \gamma_\mu n_r)(\bar n_s \gamma^\mu  n_t)$ &
&  &
$\Q_{\ell n u q}$  & $(\bar \ell^j_p  n_r)(\bar u_s q^j_t)$  \\
$\Q_{nedu}$      & $(\bar n_p \gamma_\mu e_r)(\bar d_s \gamma^\mu  u_t)$   & & &
&   \\
\hline
\multicolumn{2}{c}{$\psi^2\phi^3$} &
\multicolumn{2}{c}{$\psi^2\phi^2 D$} &
\multicolumn{2}{c}{$\psi^2 X \phi$}\\
\hline
$\Q_{n\phi}$  & $(\phi^\dagger \phi) (\bar l_p  n_r \tilde \phi)$ & $\Q_{\phi n}$   &
$i(\phi^\dagger \overset{\leftrightarrow}{D}_\mu \phi) (\bar n_p \gamma^\mu n_r) $& $\Q_{nW}$ & $(\bar \ell_p  \sigma^{\mu\nu} n_r)\tau^I \tilde \phi W_{\mu\nu}^I$ \\
&  & $\Q_{\phi ne}$   &  $i(\tilde \phi^\dagger D_\mu \phi) (\bar n_p \gamma^\mu e_r) $  & $\Q_{nB}$  & $(\bar \ell_p  \sigma^{\mu\nu} n_r)\tilde \phi B_{\mu\nu}$ \\
\hline 
\end{tabular}
\caption{\small $\mathcal{B}$- and $L$- conserving dimension-six operators in the $\nu$SMEFT in the Warsaw basis notation.}
\label{tab:SMEFTops1}
\end{subtable}

\bigskip 

\begin{minipage}{0.48\textwidth}
\begin{subtable}{\linewidth}
\centering
\begin{tabular}{c|c}
\hline 
\multicolumn{2}{c}{$\Delta \mathcal{B} = \Delta L = 1$ + h.c.}\\
\hline 
$\Q_{qqdn}$     & $\epsilon^{\alpha\beta\gamma}\epsilon_{ij}(q_p^{i\alpha\,T} C q_r^{j\beta})(d_s^{\gamma\,T}Cn_t)$ \\
$\Q_{uddn}$     & $\epsilon^{\alpha\beta\gamma}(u_p^{\alpha\,T} C d_r^\beta)(d_s^{\gamma \,T}Cn_t)$   \\
\hline  
\end{tabular}
\caption{\small $\mathcal{B}$- and $L$- violating  dimension-six operators in the $\nu$SMEFT.}
\label{tab:SMEFTops4}
\end{subtable}
\end{minipage}
\hfill 
\begin{minipage}{0.48\textwidth}
\begin{subtable}{\linewidth}
\centering
\begin{tabular}{c|c}
\hline 
\multicolumn{2}{c}{$\psi^2X$ and $\psi^2\phi^2$}\\
\hline 
$\Q_{nn B}$     & $(n^T_p C \sigma^{\mu \nu} n_r) B_{\mu \nu}$   \\
$\Q_{n n \phi \phi}$     & $(n^T_p C n_r) \phi^\dagger \phi$   \\
\hline  
\end{tabular}
\caption{\small $L$- violating dimension-five operators in the $\nu$SMEFT.}
\label{tab:SMEFTops3}
\end{subtable}
\end{minipage}
\caption{$\nu$SMEFT operator basis up to dimension-six.}
\end{table}


{\boldmath
\subsection{Choice of Flavour Basis}\label{subsec:massbasis}
}
Extending the SM with three RH neutrino fields $n_p$ (calling the resulting model to be 
$\nu$SM), the maximal flavour symmetry  group possessed by the $\nu$SM is then
\begin{equation} \label{eq:U6sym}
\rm SU(3)_q \otimes SU(3)_u \otimes SU(3)_d \otimes SU(3)_\ell \otimes SU(3)_e \otimes SU(3)_n. 
\end{equation}

\begin{table}[H]
\resizebox{\textwidth}{!}{
\renewcommand{\arraystretch}{1.5}
\begin{tabular}{l|l}
    \hline
    \toprule
\qquad {\bf Two-fermion}   &    
{\bf \qquad\qquad\qquad\qquad \qquad Four-fermion } \\  \midrule \midrule
$C^{n\phi} = U_{e_L}^\dagger C^{'n\phi} U_{n_R}$ &
    $(C^{nd})_{f_1 f_2 f_3 f_4} = (U_{n_R})_{g_2 f_2} (U_{d_R})_{g_4
      f_4} (U_{n_R})_{g_1 f_1}^* (U_{d_R})_{g_3 f_3}^* (C^{'nd})_{g_1
      g_2 g_3 g_4} $ \\[1.5mm]
$C^{\phi n} = U_{n_R}^\dagger C^{'\phi n} U_{n_R}$ &
    $(C^{nu})_{f_1 f_2 f_3 f_4} = (U_{n_R})_{g_2 f_2} (U_{u_R})_{g_4f_4} (U_{n_R})_{g_1 f_1}^* (U_{u_R})_{g_3 f_3}^* (C^{'nu})_{g_1
      g_2 g_3 g_4} $ \\[1.5mm]
$C^{\phi ne} = U_{n_R}^\dagger C^{' \phi ne} U_{e_R}$ &
    $(C^{ne})_{f_1 f_2 f_3 f_4} = (U_{n_R})_{g_2 f_2} (U_{e_R})_{g_4 f_4} (U_{n_R})_{g_1 f_1}^* (U_{e_R})_{g_3 f_3}^* (C^{'ne})_{g_1
      g_2 g_3 g_4} $ \\[1.5mm]
$C^{nW} = U_{e_L}^\dagger C^{' nW} U_{n_R}$ & 
     $(C^{nn})_{f_1 f_2 f_3 f_4} = (U_{n_R})_{g_2 f_2} (U_{n_R})_{g_4 f_4}(U_{n_R})_{g_1 f_1}^* (U_{n_R})_{g_3 f_3}^* (C^{'nn})_{g_1 g_2 g_3 g_4} $
    \\[1.5mm]
$C^{nB} = U_{e_L}^\dagger C^{'nB} U_{n_R}$ & 
      $(C^{nedu})_{f_1 f_2 f_3 f_4} = (U_{e_R})_{g_2 f_2} (U_{u_R})_{g_4 f_4} (U_{n_R})_{g_1 f_1}^* (U_{d_R})_{g_3 f_3}^* (C^{'nedu})_{g_1 
      g_2 g_3 g_4} $ \\[1.5mm]
{$C^{nnB} = U_{n_R}^T C^{'nnB} U_{n_R}$} & 
     $(C^{qn})_{f_1 f_2 f_3 f_4} = (U_{d_L})_{g_2 f_2} (U_{n_R})_{g_4 f_4} (U_{d_L})_{g_1 f_1}^* (U_{n_R})_{g_3 f_3}^* (C^{' qn})_{g_1 g_2 g_3 g_4} $
    \\[1.5mm]
{$C^{nn\phi\phi} = U_{n_R}^T C^{'nn\phi\phi} U_{n_R}$} & 
     $(C^{\ell n})_{f_1 f_2 f_3 f_4} = (U_{e_L})_{g_2 f_2} (U_{n_R})_{g_4 f_4} (U_{e_L})_{g_1 f_1}^* (U_{n_R})_{g_3 f_3}^* (C^{' \ell n})_{g_1 g_2 g_3 g_4} $
    \\[1.5mm]
& 
     $(C^{\ell n \ell e})_{f_1 f_2 f_3 f_4} = (U_{n_R})_{g_2 f_2} (U_{e_R})_{g_4 f_4} (U_{e_L})_{g_1 f_1}^* (U_{e_L})_{g_3 f_3}^* (C^{' \ell n \ell e })_{g_1
      g_2 g_3 g_4} $ \\[1.5mm]
& 
     $(C^{\ell n q d (1)})_{f_1 f_2 f_3 f_4} = (U_{n_R})_{g_2 f_2} (U_{d_R})_{g_4 f_4} (U_{e_L})_{g_1 f_1}^* (U_{d_L})_{g_3 f_3}^* (C^{' \ell n qd (1)})_{g_1 g_2
      g_3 g_4} $ \\[1.5mm]
& 
    $(C^{\ell n q d (3)})_{f_1 f_2 f_3 f_4} = (U_{n_R})_{g_2 f_2} (U_{d_R})_{g_4 f_4} (U_{e_L})_{g_1 f_1}^* (U_{d_L})_{g_3 f_3}^* (C^{'\ell n qd  (3)})_{g_1
      g_2 g_3 g_4} $ \\[1.5mm]
  & 
   $(C^{\ell n u q })_{f_1 f_2 f_3 f_4} = (U_{n_R})_{g_2 f_2} (U_{d_L})_{g_4 f_4} (U_{e_L})_{g_1 f_1}^* (U_{u_R})_{g_3 f_3}^* (C^{'\ell n u q})_{g_1
      g_2 g_3 g_4} $ \\[1.5mm]
        & 
   {$(C^{qqdn})_{f_1 f_2 f_3 f_4} = (U_{d_L})_{g_2 f_2} (U_{n_R})_{g_4 f_4} (U_{d_L})_{g_1 f_1}^* (U_{d_R})_{g_3 f_3}^* (C^{'qqdn})_{g_1
      g_2 g_3 g_4} $} \\[1.5mm]
        & 
   {$(C^{uddn})_{f_1 f_2 f_3 f_4} = (U_{d_R})_{g_2 f_2} (U_{n_R})_{g_4 f_4} (U_{u_R})_{g_1 f_1}^* (U_{d_R})_{g_3 f_3}^* (C^{'uddn})_{g_1
      g_2 g_3 g_4} $} \\[1.5mm]
\bottomrule
\end{tabular}}
		\bigskip
\caption{ Definitions (preserving ${ SU(3)^6}$) for the $\nu$SMEFT WCs of operators involving fermions 
in the down-basis (i.e. obtained by setting $U_q = U_{d_L},~ U_\ell = U_{e_L},~ U_u = U_{u_R},~ U_d = U_{d_R}, ~U_e = U_{e_R}$ and $U_n = U_{n_R}$). 
\label{tab:wcsrot}}
\end{table}
\noindent
The flavour symmetry transformations can be defined through:
\begin{align}
\begin{aligned}
q &\to  U_q q\,,  \quad\quad\, \ell \to  U_\ell \ell\,,  \\
u &\to  U_u u\,,  \quad\quad d \to  U_d d\,,  \\
e &\to  U_e e\,,  \quad\quad\, n \to  U_n n \,,
\end{aligned}
\end{align}
with $U_\psi$ to be unitary matrices. The $\rm SU(3)_n$ flavour symmetry can be broken down to $\rm O(3)_n$ by the Majorana mass term for the neutrinos  (see \eqref{eq:nuSMEFTLag}).  The remaining symmetry of $\nu$SM is broken by the Yukawas terms \cite{Cirigliano:2005ck}. In the absence of Majorana mass term, $\nu$SMEFT also possess 
the full flavour symmetry \eqref{eq:U6sym} up to redefinitions of the WCs. 
 Thus the flavour basis of $\nu$SMEFT is not unique.

Two convenient choices 
of bases are defined by assigning specific values to $U_q$ and $U_\ell$: 
\begin{align}
U_q &= U_{d_L}\,, \quad   \quad U_\ell = U_{e_L} \quad  \quad \textit{\rm (down-basis)}\,, \\
U_q &= U_{u_L} \,, \quad \quad  U_\ell = U_{\nu_L} \quad \quad ~ \textit{\rm (up-basis)}. 
\end{align}
In the down-basis (up-basis), the down-type (up-type) Yukawa matrices take a diagonal form. 
In Table.~\ref{tab:wcsrot}, we show the redefinitions of the $\nu$SMEFT WCs in the down-basis 
convention. The corresponding redefinitions for the pure SMEFT operators can be found in 
Ref.~\cite{Dedes:2017zog}. As a result in these two bases, the only unknown parameters
are the WCs, the SM parameters, and $\hat Y_n$.  
In the current implementation of the $\nu$SMEFT in \wilson{} 
we have adopted the down-basis convention assuming no Majorana mass term.

\section{Evolution in the $\nu$SMEFT}\label{sec:evolution}
The RG running in the $\nu$SMEFT from the NP scale to the EW scale is controlled by:
\be
\dot C_i(\mu) = {16\pi^2} \mu {d \over d\mu} C_i(\mu)   = \hat \gamma_{ij}(g_1, g_2, g_3, \hat Y_\psi)  C_j(\mu)\,, 
\ee
where $\mu$ is the renormalization scale, and $g_i$ and $\hat Y_\psi$ are the gauge couplings and Yukawa matrices for the quarks, leptons, and sterile neutrinos. 
The latter gives rise to a neutrino Dirac mass term after EW symmetry breaking (EWSB). We implemented the full gauge and Yukawa dependence of the ADM $\hat \gamma_{ij}$. For the code implementation, the explicit expressions for $\hat \gamma_{ij}$ are taken from \cite{Datta:2020ocb} (gauge-coupling dependence) and \cite{Datta:2021akg} (Yukawa dependence)\footnote{While this work was in preparation, in a recent study \cite{Ardu:2024tzb} the missing Yukawa terms in the ADMs have been computed. These will be included in the future update of the \wilson{} package.}. The running of the Baryon number violating operators was taken from \cite{Alonso:2014zka}. For the purpose of implementation in \wilson{}, to match the \texttt{WCxf} convention, we have used the conjugate of Yukawas (denoted by $\hat Y_\psi$) as compared to Eq.~\eqref{eq:Yuklag} (corresponding to the original convention of Ref.~\cite{Datta:2021akg}). The structure of the ADMs within the $\nu$SMEFT due to gauge couplings exhibits a block structure for the SMEFT and $\nu$SMEFT specific operators, meaning the corresponding operators mix only among themselves. However, the Yukawa-dependent ADMs (or the ADMs depending upon the combination of Yukawa and gauge couplings) also introduce mixing between pure $\nu$SMEFT and pure SMEFT operators.

For the purpose of RG evolution within the $\nu$SMEFT, all dimension-four parameters are required at the input scale. In contrast to the SMEFT case, where all dimension-four parameters at the NP scale can be determined from their corresponding SM values at the EW scale, the neutrino Yukawa couplings $\hat Y_n$ in the $\nu$SMEFT  belong to the unknown NP parameter category, which must be inputted along with the $\nu$SMEFT WCs to solve the above RGEs. At dimension-six level $\hat Y_n$ also receives corrections from the $\wc[]{n\phi}{}$ WC. Also, one can add a dimension-four Majorana mass term, in addition to the dimension-five part originating within SMEFT from the Weinberg operator. But such terms do not affect the $\nu$SMEFT ADMs directly \cite{Datta:2020ocb,Datta:2022czw, Fuyuto:2024oii}, in most cases.

\section{$\nu$SMEFT implementation in \wilson{}}\label{sec:nusmeft_wilson}
The inclusion of $\nu$SMEFT evolution in \wilson{} has the following components:
\begin{enumerate}
\item Addition of the $\nu$SMEFT as a new EFT in the \WCxf{} format.
\item Addition of a basis file for the $\nu$SMEFT in the  \WCxf{} format. In \wilson{} we continue to call this the {\tt Warsaw} basis,  as this basis has been inspired by the corresponding SMEFT Warsaw basis. 
\item Addition of the $\nu$SMEFT ADMs in \wilson{}.
\end{enumerate}
Apart from that, we have made important changes to the \wilson{} package. The major one is replacing the \texttt{SMEFT} class with the \texttt{EFTEvolve} class. While the former was dedicated to the SMEFT evolution, the latter is designed to be able to perform the RG evolution within all three EFTs: SMEFT, $\nu$SMEFT, and WET. In the current version, we do not utilize it for the WET evolution, which 
is kept as it was. In the forthcoming versions, we plan to use \texttt{EFTEvolve} for the WET as well. The main difference between the two classes is that the \texttt{\_\_init\_\_} method of the \texttt{EFTEvolve} class also requires the \textit{beta-function} in form of a dictionary, in addition to the \WCxf{} instance representing a parameter point in the EFT space. At the user level, the evolution within the $\nu$SMEFT can be performed using a few lines of code:
\begin{lstlisting}[language=iPython]
from wilson import Wilson
mywilson = Wilson({'nd_1111': 1e-6, 'lnle_1111':1e-6},
                  scale=1e3, eft='nuSMEFT', basis='Warsaw')
mywilson.match_run(91, 'nuSMEFT', 'Warsaw')                  
\end{lstlisting}
where the values of input dimension-six WCs need to be specified in units of ${\rm GeV}^{-2}$.

\par 
Unlike in the SMEFT, the dimension-four parameter $\hat Y_n$ has to be provided as an input parameter at the input scale. This can be done by the command \texttt{set\_option}, with the option \texttt{yukawa\_scale\_in}, which allows the user to input any Yukawa matrix ($\hat{Y}_\psi, \; { \psi = u, d, e, n} )$. Its usage is demonstrated below:
\begin{lstlisting}[language=iPython]
from wilson import Wilson
import numpy
mywilson = Wilson({'nd_1111': 1e-6, 'lnle_1111':1e-6},
                  scale=1e3, eft='nuSMEFT', basis='Warsaw')
mywilson.set_option('yukawa_scale_in', {'Gn': numpy.eye(3), 'Gu': numpy.zeros((3,3))})       
\end{lstlisting}
The Yukawa matrices (represented by ${\tt G} \psi$ within the code) are given in a Python dictionary, with the key being the name and its values passed on as a {\tt numpy} array. In this example, as an illustration, we input only two Yukawa matrices, where we set $\hat{Y}_n$ as a $3\times3$ matrix with diagonal entries equal to 1, and $\hat{Y}_u$ is a $3\times3$ null matrix.
If not specified, the SM Yukawas $\hat Y_u, Y_d, Y_e$ are internally determined within {\tt wilson}, while $\hat Y_n$ 
is set to zero. 

\par 
Similarly, another option named \texttt{gauge\_higgs\_scale\_in} for the 
command \texttt{set\_option} allows the user to enter gauge couplings $g_1, g_2$ and $g_3$ (denoted by \texttt{gp, g} and \texttt{gs}, respectively in the program), and Higgs parameters $m_H^2$ and $\lambda$ (denoted by \texttt{m2} and \texttt{Lambda}, respectively in the program) in the form of a dictionary. Its usage is demonstrated in the following code snippet:
\begin{lstlisting}[language=iPython]
from wilson import Wilson
import numpy
mywilson = Wilson({'nd_1111': 1e-6, 'lnle_1111':1e-6},
                  scale=1e3, eft='nuSMEFT', basis='Warsaw')
mywilson.set_option('gauge_higgs_scale_in', {'g': 0, 'gp': 1, 'gs': 0.5, 'Lambda': 100, 'm2': 100})       
\end{lstlisting}
The values of $m_H^2$ and $\Lambda$ have to be given in the units of \textit{$\rm GeV^2$} and GeV, respectively.

\section{Comparison and cross-checks}\label{sec:comparison}
To test the proper functioning of the new code, we made several checks by running internal test functions, as well as comparing the output with known results. Two major checks of the output were performed for the case of the SMEFT and the $\nu$SMEFT, as only the corresponding part of the code was upgraded (the WET EFT remains unmodified). 
\subsection{SMEFT evolution}
To test the proper functioning of our upgraded code for the SMEFT, we compare the results from our new version of the code with the previous version. For this purpose {dictionaries} of all 1635 SMEFT Wilson coefficients in the Warsaw basis (including all the non-redundant combinations of the indices) with randomly generated input values at $\Lambda = $ 1 TeV {were} generated. The running is then performed to the EW scale using the new and the old versions of the implementation. As expected, the output matched precisely, confirming the stable working of the new code for the SMEFT case.

\subsection{$\nu$SMEFT evolution}
To verify the results of the novel implementation for the $\nu$SMEFT, we have reproduced the results of the article \cite{Datta:2020ocb}. For this test, we set the indices $prst = 1111$ and list the $16 \times 16$ $\nu$SMEFT WCs in the basis 
\beqa
\vec{\mathbf{\C}} &=&\{\C_{nd},\,\C_{nu},\,\C_{ne},\,\C_{qn},\,\C_{\ell n},\,\C_{\phi n},\,\C_{n\phi},\,\C_{nW},\,\C_{nB},\,C^{(1)}_{\ell nqd},\,C^{(3)}_{\ell nqd},\,\C_{nedu},\,\C_{\ell n \ell e},\,\C_{\ell nuq},\,\C_{\phi ne},\,\C_{nn}\}\,.\notag\\
	\eeqa
The 16 WCs at the EW scale and at $\Lambda$ = 1 TeV are then related by 
\beq
\frac{\delta \mathcal{C}(M_Z)}{10^{-3}}	= 
\begin{psmallmatrix}
-0.89 & 1.77 & -0.89 & 0.89 & -0.89 & 0.44  & & & & & & & & & & \\
1.77 & -3.54 & 1.77 & -1.77 & 1.77 & -0.89 &  & & & & & & & & & \\ 
-2.66 & 5.32 & -2.66 & 2.66 & - 2.66 & 1.33 & & & & & & & & & & \\ 
0.44 & -0.89 & 0.44 & -0.44 & 0.44 & -0.22 & & & & & & & &  & & \\ 
-1.33 & 2.66 & -1.33 & 1.33 & -1.33 & 0.66 & & & & & & & & & & \\ 
1.33 & -2.66 & 1.33 & -1.33 & 1.33 & -0.66 & & & & & & & & & & \\ 
& & & & &  & 46.26 & 39.39 & -8.9 & & & & & & & & \\ 
& & & & &  & 0 & 7.17 & 5.27 & & & & & &  &\\ 
& & & & &  & 0 & 15.81 & -1.19 & & & & & &  &\\ 
& & & & &  &  & &  & 136.72 & -107.42 & & & &  &\\ 
& & & & &  &  & & & -2.24 & -25.01 & & & &  &\\
& & & & &  &  & & &  & & 7.97 & & &  &\\
& & & & &  &  & & &  & & & -0.31 & &  &\\
& & & & &  &  & & &  & & &  & 138.71 &  &\\
& & & & &  &  & & &  & & &  & & 5.98 &\\
& & & & &  &  & & &  & & &  & & & 0\\
\end{psmallmatrix}
\mathcal{C}(\Lambda)\,,
\eeq
The results shown in this matrix agree well with those given in \cite{Datta:2020ocb}. The running effects in the $6 \times 6$ and $3 \times 3$ blocks are small because only EW gauge couplings contribute. The mixing in the $2 \times 2$ block is large as it is governed by QCD running.

\section{Summary}\label{sec:summary}

The lack of observations of new particles at the LHC indicates the scale of NP to be well above the EW scale. Potential BSM effects can then be encoded in terms of Wilson coefficients of the SMEFT in a general manner. However, the field content of the SMEFT is restricted to SM particles, which makes it unfit for certain NP models containing light sterile neutrino states. Adding such fields to the SMEFT results in new higher-dimensional operators starting at the dimension-five and dimension-six levels. Those operators can mix within themselves as well as with the standard SMEFT operators at the one-loop level. While anomalous dimensions for such mixing terms are known, including their effects in the physical observables requires solving the corresponding RGEs. For this purpose, a careful consideration of properly evaluating the SM parameters at the NP scale is a must, but often ignored in phenomenological studies, where typically only first leading log solutions are considered. 
\par
In this work, we have presented an upgrade of the \wilson{} package, which is the first public code that includes the full numerical evolution of the $\nu$SMEFT parameters including subtle effects such as back-rotation of flavour bases. For the purpose of RG running within the $\nu$SMEFT, the neutrino Yukawa matrix has to be provided along with the Wilson coefficients while all other dimension-four parameters are internally evaluated within \wilson{}, unless provided by the user. 
This implementation allows to include RG effects in studies related to light sterile neutrino particles and their correlations with other sectors such as flavour physics. In the future, we plan to include the WET augmented with sterile neutrinos in \wilson{}, together with the corresponding matching conditions from the $\nu$SMEFT and a proper treatment of neutrino masses and mixing.

\begin{acknowledgments}

We thank E. Mereghetti for the useful discussions.
The work of J.A. is supported by the European Union’s Horizon 2020 research and innovation program under the Marie Sk\l{}odowska-Curie grant agreement No.~101145975 - EFT-NLO.
The work of J.K. was supported by the US Department of Energy Office 
and by the Laboratory Directed Research and Development (LDRD) program of Los Alamos National Laboratory under project numbers
20230047DR, 20220706PRD1.
Los Alamos National Laboratory is operated by Triad
National Security, LLC, for the National Nuclear Security Administration of the U.S. Department of Energy
(Contract No. 89233218CNA000001).

\end{acknowledgments}

\appendix

\addcontentsline{toc}{section}{References}
\small
\bibliographystyle{JHEP}
\bibliography{nusmeft}

\end{document}